\documentclass[prl,twocolumn,showpacs,amsmath,amssymb,superscriptaddress]{revtex4-2}
\usepackage{bm} 
\usepackage{dcolumn} 
\usepackage{graphicx} 
\usepackage{amsmath} 
\usepackage{amssymb} 
\usepackage{amsfonts} 

\begin{document}
\title{Fermi surface topology and renormalization of bare ellipticity in interacting anisotropic electron gas }

\author{Seongjin Ahn}
\affiliation{Condensed Matter Theory Center and Joint Quantum Institute, Department of Physics, University of Maryland, College Park, Maryland 20742-4111, USA}
\author{S. Das Sarma}
\affiliation{Condensed Matter Theory Center and Joint Quantum Institute, Department of Physics, University of Maryland, College Park, Maryland 20742-4111, USA}

\date{\today}

\begin{abstract}
We investigate effects of electron-electron interactions on the shape of the Fermi surface in an anisotropic two-dimensional electron gas using the `RPA-$GW$' self-energy approximation. We find that the interacting Fermi surface deviates from an ellipse, but not in an arbitrary way. The interacting Fermi surface has only two qualitatively distinct shapes for most values of $r_s$. The Fermi surface undergoes two distinct transitions between these two shapes as $r_s$ increases. For larger $r_s$, the degree of the deviation from an ellipse rapidly increases, but, in general, our theory provides a justification for the widely used elliptical Fermi surface approximation even for the interacting system since the non-elliptic corrections are quantitatively rather small except for very large $r_s$.

\end{abstract}

\maketitle

{\em Introduction.}--- 
The concept of a Fermi surface is one of the great triumphs of quantum physics and is a central paradigm in solid state physics with the physics of all metallic systems being closely tied to their Fermi surface properties.  In particular, the topology of a Fermi surface plays a vital role in determining low-energy physical properties 
of metals. This naturally raises a question as to how the Fermi surface {\it shape} evolves under the influence of electron-electron interactions. The answer is obvious for an isotropic system because the rotational symmetry forces the Fermi surface to be a circle. 
When the rotational symmetry is explicitly broken in the noninteracting system by having an elliptic Fermi surface with two different effective masses, however, the Fermi surface is not necessarily constrained to any specific shape in the corresponding interacting system. Since the Fermi surface is anisotropic for most realistic materials because of lattice-induced band structure effects, there has been interest in determining the shape of the interacting Fermi surface in an anisotropic system \cite{Wu1995,Roldan2006, Tolsma2016,Krishna2019,Leaw2019}. In the current work, we investigate how interactions affect the low-energy properties of an anisotropic electron gas by directly calculating the self-energy including the full dynamical effects of the anisotropy. Our starting point is a noninteracting anisotropic two dimensional electron gas characterized by two unequal effective masses, i.e., 
\begin{equation}
\varepsilon(\bm k)=\frac{k_x^2}{2m_\mathrm{H}}+\frac{k_y^2}{2m_\mathrm{L}},
\end{equation}
as for example, in Si 110 inversion layers and other semiconductor structures \cite{Ando1982}.

\begin{figure}[!htb]
    \centering
    \includegraphics[width=1\linewidth]{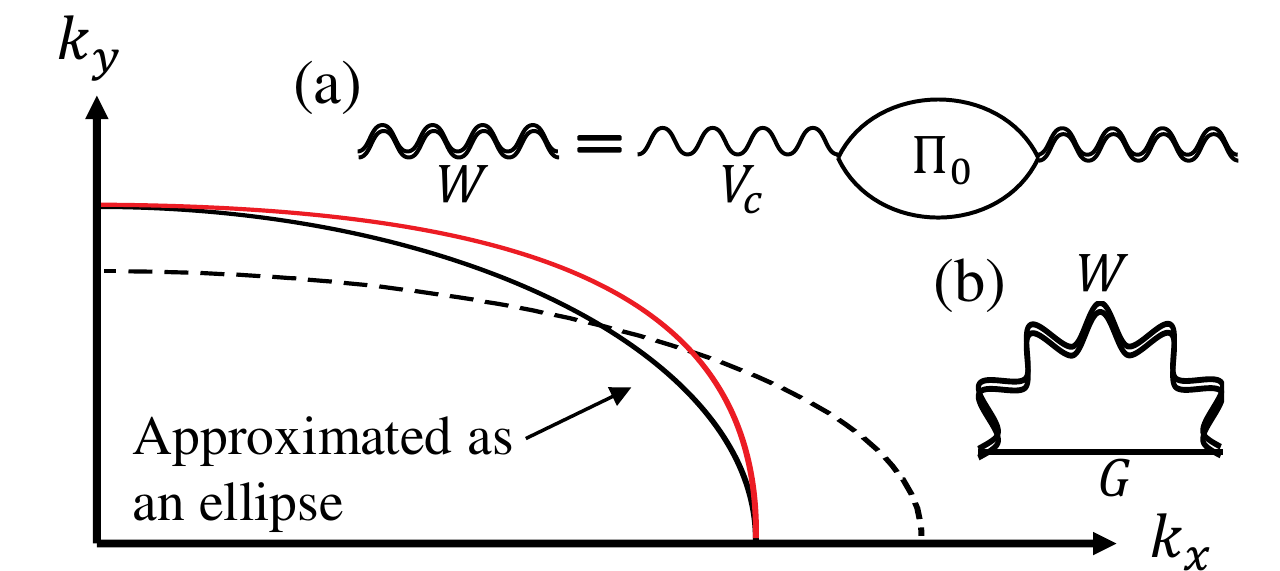}
    \caption{Schematics of noninteracting (black-dashed) and interacting Fermi surfaces (solid). The top-solid (red) line represents the interacting non-elliptic Fermi surface whereas the bottom-solid (black) line shows the widely used elliptic approximation. (a) Series of diagrams corresponding to RPA. The wiggly line represents the Coulomb interaction and $\Pi_0$ the bare polarizability (b) Feynman-Diagram representing the self-energy within RPA-$GW$ approximation, where $W$ refers to a dynamically screened Coulomb interactions within the RPA. 
      }
    \label{fig:FS_schematic}
\end{figure}

In spite of the manifest conceptual and practical importance of the question (i.e. ``What is the shape of the interacting Fermi surface when the noninteracting Fermi surface is an ellipse?") in semiconductors with anisotropic band mass, there have been few attempts to address the question in spite of its importance already being obvious 50 years ago-- typically, an effective isotropic approximation has invariably been used, often using the corresponding density of states effective mass, in calculating the interaction effects \cite{Luttinger1955, Brinkman1972, Brinkman1972a, Combescot1972, Beni1978}. A recent study approached this question by approximating the interacting Fermi surface by an ellipse (see black solid line in Fig.~\ref{fig:FS_schematic}) \cite{Tolsma2016}.
To our knowledge, however, there exists no complete study of the topology of the interacting anisotropic Fermi surface beyond the elliptical shape approximation.

In this work, we investigate the deviation of the renormalized anisotropic Fermi surface from an ellipse by calculating the self-energy within the leading-order dynamically-screened Coulomb interaction (RPA) or the $GW$ approximation \cite{quinn1958electron, Hedin1965}, using the full band anisotropy in the self-energy calculation. We first show that the renormalized effective masses $m_\mathrm{H}^*$ and $m_\mathrm{L}^*$ vary as a function of the location on the Fermi surface (and hence, the interacting Fermi surface is no longer elliptic), and analyze their behaviors in several regimes of the dimensionless Coulomb interaction strength parameter $r_s$ \cite{Mahan2000}. We then demonstrate the evolution of the Fermi surface by using the obtained results for the effective mass. We find that there are two typical shapes for the interacting Fermi surface defined by the effective mass, and the degree of the deviation from the noninteracting bare elliptic shape becomes significant for strongly interacting systems at large $r_s$ (although the interacting Fermi surface is nonelliptic for all $r_s$). Although we find that the bare elliptic Fermi surface is destroyed, in principle, by interaction-induced non-elliptic effects, the magnitude of this nonellipticity is rather small in general, and hence our work provides a justification for the extensively used elliptic approximation for interacting anisotropic systems. In fact, we show in details how interaction actually strongly suppresses the ellipticity while at the same time introducing small nonelliptic effects..

{\em Theory.}---  The self-energy for an anisotropic electron gas within the RPA-$GW$ approximation is given by 
 [Fig.~\ref{fig:FS_schematic}(b)],
\begin{align}
\Sigma({\bm k},i\omega_n)\!=&-\!\int\!\frac{d^2 q}{(2\pi)^2}\frac{1}{\beta}\sum_{i\Omega_n}W(\bm q,i\Omega_n) \nonumber \\ 
&\times G_0(\bm k + \bm q, i\omega_n+i\Omega_n),
\end{align}
where $G_0=\left(i\omega_n+i\Omega_n-\xi_{\bm k + \bm q} \right)^{-1}$ is the bare Green's function, $\omega_n$ and $\Omega_n$ are Matsubara frequencies, $\beta=(k_\mathrm{B}T)^{-1}$, $T$ is the temperature, $k_\mathrm{B}$ is the Boltzmann constant, $\xi_{\bm k}=\varepsilon_{\bm k} - \mu$, and $\mu$ is the chemical potential. Here $W(\bm q,i\Omega_n)=v_c(\bm q)/\varepsilon(q,i\Omega_n)$ is the dynamically screened Coulomb interaction where $v_c(\bm q)=2\pi e^2/|\bm q|$ is the two-dimensional (2D) bare Coulomb interaction and $\varepsilon(q,i\Omega_n)=1-v_c(\bm q)\Pi_0(\bm q,i\Omega_n)$ is the two-dimensional dielectric function obtained within RPA [Fig.~\ref{fig:FS_schematic}(a)] with $\Pi_0(\bm q,\omega)$ being the noninteracting polarization function for an anisotropic two-dimensional electron gas. $\Pi_0(\bm q,\omega)$ can be obtained exactly from the existing result \cite{Stern1967} for an ideal isotropic gas by rescaling $m_e\rightarrow m_\mathrm{DOS}$, $k_x\rightarrow \sqrt{\frac{m_\mathrm{DOS}}{m_\mathrm{H}}}k_x$, and $k_y\rightarrow  \sqrt{\frac{m_\mathrm{DOS}}{m_\mathrm{L}}}k_y$, where $m_e$ refers to the bare electron mass, and $m_\mathrm{DOS}=(m_\mathrm{H}m_\mathrm{L})^{1/2}$ as arising in the definition of the density of states for the noninteracting system. 

It is useful to divide the $GW$ self-energy into the exchange and correlation parts: $\Sigma=\Sigma^\mathrm{ex}+\Sigma^\mathrm{corr}$. The exchange part corresponds to the self-energy with bare Coulomb interaction. The zero-temperature exchange self-energy is given by
\begin{align}
\Sigma^{\mathrm{ex}} ({\bm k})\!=&-\!\int\!\frac{d^2 q}{(2\pi)^2} \Theta(-\xi_{\bm k+\bm q}) v_c(\bm q).
\label{eq:self_energy_ex}
\end{align}
The correlation part contains all contributions beyond bare exchange interaction, and is written as
\begin{align}
    \Sigma^\mathrm{corr}({\bm k},i\omega_n)\!=&-\!\int\!\frac{d^2 q}{(2\pi)^2}\frac{1}{\beta}\sum_{i\Omega_n}
    \left[\frac{1}{\varepsilon(\bm q,i \Omega_n)}-1\right]
    \nonumber \\ 
&\times G_0(\bm k + \bm q, i\omega_n+i\Omega_n).    
\end{align}
We express the retarded correlation self-energy as a sum of two terms \cite{quinn1958electron}: 
$\Sigma^\mathrm{corr}=\Sigma^\mathrm{line}+\Sigma^\mathrm{res}$.
The line part $\Sigma^\mathrm{line}$ is obtained by performing an analytic continuation, i.e., $i\omega_n\rightarrow\omega+i\eta$. Using $\beta^{-1}\sum_{i\Omega_n}=\int_{-\infty}^{\infty}d\Omega$ in the $T\rightarrow0$ limit, we derive 
\begin{align}
\Sigma^{\mathrm{line}} ({\bm k},\omega)\!=&-\!\int\!\frac{d^2 q}{(2\pi)^2}\int_{-\infty}^{\infty}\!\frac{d\Omega}{2\pi} 
\frac{v_c(\bm q)}{\xi_{\bm k+\bm q}-\omega-i\Omega} \nonumber \\ 
&\times\left[\frac{1}{\varepsilon(\bm q,i \Omega)}-1\right].
\label{eq:self_energy_line}
\end{align}
Since the Matsubara summation should be done before analytic continuations, $\Sigma^{\mathrm{line}}$ is not the entire correlation self-energy. $\Sigma^{\mathrm{res}}$ is the difference arising from exchanging the order of the Matsubara frequency summation and analytic continuations, and is given by 
\begin{align}
\Sigma^{\mathrm{res}} ({\bm k},\omega)\!=&\!\int\!\frac{d^2 q}{(2\pi)^2} \left [\Theta(\omega-\xi_{\bm k+\bm q}) - \Theta(-\xi_{\bm k+\bm q}) \right ] \nonumber \\ 
&\times v_c(\bm q)\left[\frac{1}{\varepsilon(\bm q,\xi_{\bm k+\bm q}-\omega)}-1\right].
\label{eq:self_energy_res}
\end{align}

{\em Effective Mass.}--- Once we know the self-energy, we can calculate the renormalized single particle energies by solving the Dyson's equation
\begin{equation}
E(\bm k)=\varepsilon(\bm k)+\mathrm{Re}\left[\Sigma(\bm k,\omega)|_{\omega=E(\bm k)-\mu}\right].
\label{eq:Dyson}
\end{equation}
Within the on-shell approximation which in this context is the first iteration of Dyson's equation, the self-energy is evaluated only at $\omega=\varepsilon(\bm k)-\mu$, yielding 
\begin{equation}
E({\bm k})=\varepsilon(\bm k)+\mathrm{Re}\left[\Sigma(\bm k,\varepsilon(\bm k)-\mu)\right].
\label{eq:onshell_energy}
\end{equation}
We assume that the renormalized energy dispersion is written in a form such as 
\begin{equation}
    E(\bm k)=E(0)+\frac{k_x^2}{2m^*_x(\bm k)}+\frac{k_y^2}{2m^*_y(\bm k)},
\label{eq:renorm_energy}
\end{equation}
where the momentum-dependent masses $m^*_x(\bm k)$ and $m^*_y(\bm k)$ absorb any terms that distort the Fermi surface from an ellipse. 
Note that the momentum dependence disappears in the limit where the renormalized Fermi surface is a perfect ellipse. 
Assuming that $m^*_i(\bm k)$ varies slower than $k_i^2$ on the Fermi surface, where $i=x,y$, the renormalized energy dispersion expanded around the Fermi surface is given by
\begin{equation}
    E(\bm k)\approx\mu^*+(k_x-k_{\mathrm{F}_x})\frac{k_{\mathrm{F}_x}}{m^*_x(\bm k_\mathrm{F})} + (k_y-k_{F_y})\frac{k_{F_y}}{m^*_y(\bm k_\mathrm{F})},
\label{eq:renorm_energy_expanded}
\end{equation}
where $\mu^*$ is the renormalized Fermi energy. Using Eq.~(\ref{eq:renorm_energy_expanded}), we can find an expression for the renormalized effective mass dependent on the location on the Fermi surface:
\begin{equation}
m^*_i(\bm k_\mathrm{F})=k_i[\partial E(\bm k)/\partial k_i]^{-1}|_{\bm k = \bm k_\mathrm{F}},
\label{eq:effective_mass_formula}
\end{equation}
which is a generalization of the standard definition of the effective mass that takes into account its momentum-dependent nature. Our interacting Fermi surface is defined by Eq.~(\ref{eq:effective_mass_formula}). By taking the derivative of Eq.~(\ref{eq:onshell_energy}), we obtain the renormalized effective mass to be
\begin{align}
    \frac{m_i^*(\bm k_\mathrm{F})}{m_i}=\left\{1+\frac{m_i}{k_i}\frac{\partial \mathrm{Re}\left[\Sigma(\bm k,\xi_{\bm k})\right]}{\partial k_i}\Bigr|_{\substack{\bm k=\bm k_\mathrm{F}}}\right\}^{-1}.
\end{align}
In the following, we present results of the calculated effective mass for an anisotropic 2D electron gas. Throughout the paper, we set $m_x\rightarrow m_\mathrm{H}=10m_e$ and $m_y\rightarrow m_\mathrm{L}=m_e$. 

\begin{figure}[!htb]
    \centering
    \includegraphics[width=1\linewidth]{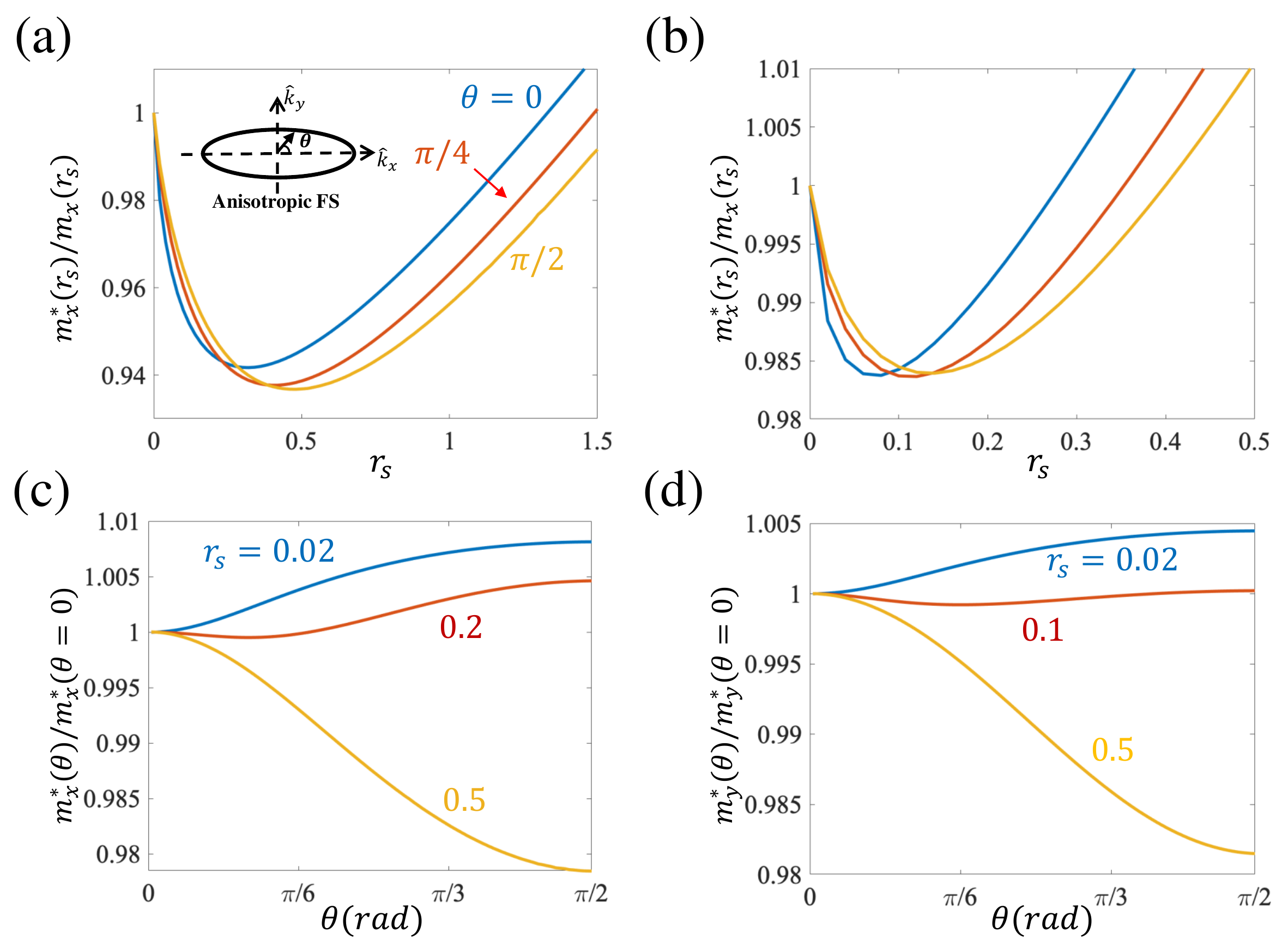}
    \caption{(a),(b) Numerically calculated effective mass as a function of $r_s$ at $\theta=0$, $\pi/4$, and $\pi/2$, where $\theta$ is the angle from the semi-major axis [see the inset in (a)]. Black-dotted boxes indicate regions where the effective mass curves cross each other (c),(d) Plots of the effective mass as a function of $\theta$ at $r_s=0.02$, $0.2$ and $0.5$. Each plot is normalized by $m^*(\theta=0)$ which refers to the effective mass at $\theta=0$. 
    }
    \label{fig:effective_mass_small_rs}
\end{figure}

Figures~\ref{fig:effective_mass_small_rs}(a) and \ref{fig:effective_mass_small_rs}(b) present calculated $m_\mathrm{H}(r_s)$ and $m_\mathrm{L}(r_s)$ for small $r_s$ at $\theta=0$, $\pi/4$ and $\pi/2$. It should be noted that $r_s^\mathrm{min}(\theta)$, which we define to be the value of $r_s$ where the effective mass at an angle of $\theta$ is minimum, is shifted to the right with increasing $\theta$. This leads to a crossover of effective mass curves [see black-dotted box in Figs.~\ref{fig:effective_mass_small_rs}(a) and \ref{fig:effective_mass_small_rs}(b)], having a direct consequence on the angular behavior of the effective mass. Figs.~\ref{fig:effective_mass_small_rs}(c) and \ref{fig:effective_mass_small_rs}(d) show the effective mass as a function of $\theta$ for a fixed $r_s$. Before the crossover occurs, the effective mass monotonically increases with increasing $\theta$ [$r_s=0.02$ in Figs.~\ref{fig:effective_mass_small_rs}(c) and \ref{fig:effective_mass_small_rs}(d)].
As $r_s$ increases up to the crossover regime, the effective mass at small $\theta$ starts decreasing but it keeps its increasing behavior at large $\theta$, resulting in a local minimum at an arbitrary $\theta$ off the symmetry axes [$r_s=0.2$ and $0.1$ in Figs.~\ref{fig:effective_mass_small_rs}(c) and \ref{fig:effective_mass_small_rs}(d), respectively]. 
As $r_s$ increases further, the local minimum is shifted to a larger $\theta$ expanding the region where the effective mass decreases. When $r_s$ is large enough to be out of the crossover regime, the local minimum disappears and the effective mass shows a decreasing behavior over the whole range of $\theta$ [$r_s=0.5$ in Figs.~\ref{fig:effective_mass_small_rs}(c) and \ref{fig:effective_mass_small_rs}(d)]. The results of Fig.~\ref{fig:effective_mass_small_rs} are restricted to $r_s<1$, where our RPA theory is essentially exact because of the weak-coupling nature of the system, but the qualitative distortion of the interacting Fermi surface topology is already apparent. 

\begin{figure}[!htb]
    \centering
    \includegraphics[width=1\linewidth]{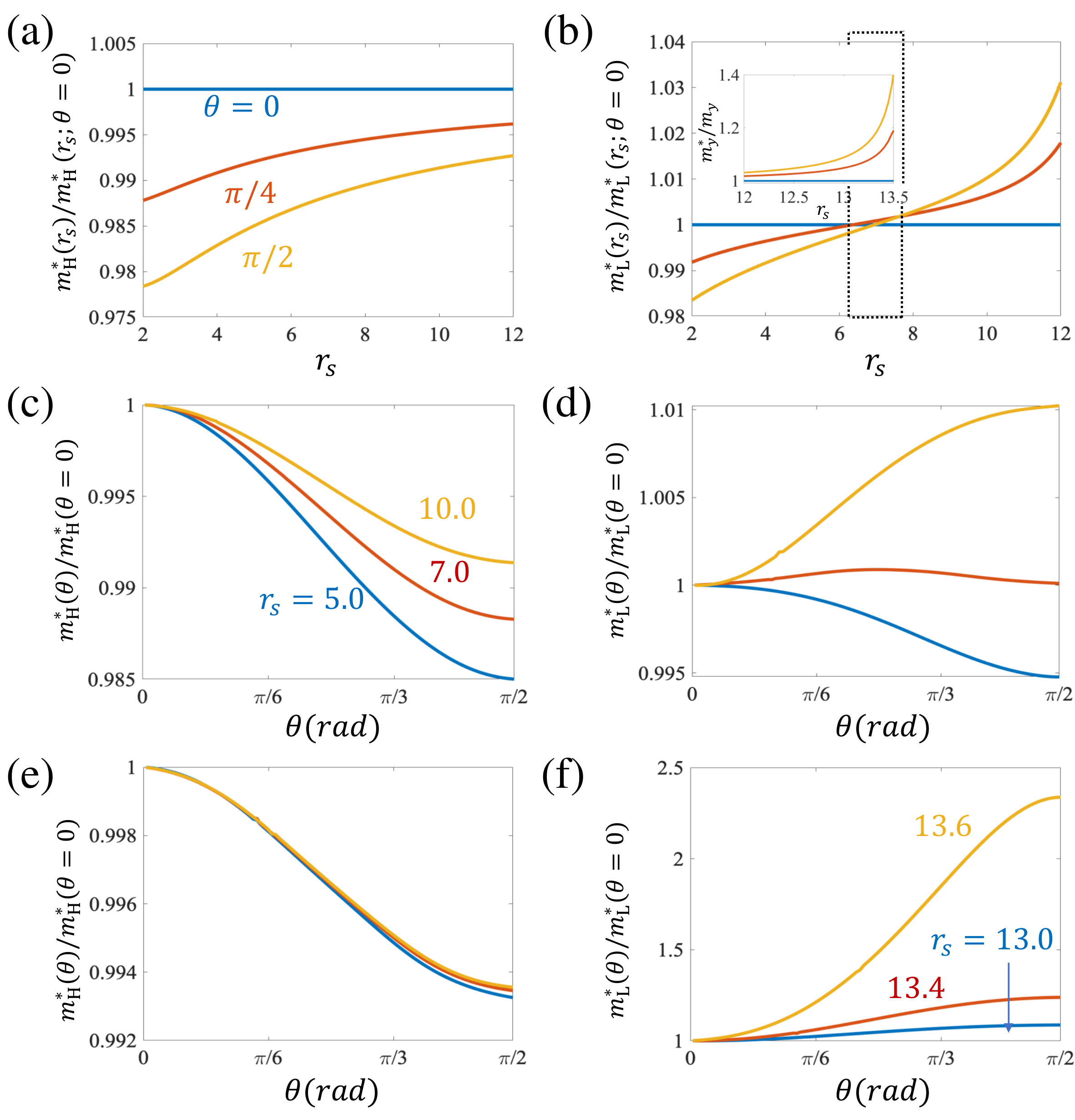}
    \caption{(a),(b) Numerically calculated effective mass as a function of $r_s$ at $\theta=0$, $\pi/4$, and $\pi/2$.
    Here each plot is normalized by the result at $\theta=0$, i.e., $m^*\left(\theta=0\right)$. The inset in (b) shows results for larger $r_s>12$, where the effective mass increases more rapidly. The black-dotted box in (b) indicates the region where the effective mass curves pass each other. (c),(d) Plots of the effective mass as a function of $\theta$ for $r_s=5.0$, $7.0$, $10.0$ and (e),(f) for still larger $r_s$. 
    }
    \label{fig:effective_mass_large_rs}
\end{figure}

In Figs.~\ref{fig:effective_mass_large_rs}(a) and \ref{fig:effective_mass_large_rs}(b), we present the calculated effective mass for large $r_s>2$ at $\theta=0$, $\pi/4$ and $\pi/2$. We normalize the result at each angle by the result at $\theta=0$ for a clear distinction between the plots.
Note that in Fig.~\ref{fig:effective_mass_large_rs}(a) the difference between the effective mass curves for different angles becomes smaller as $r_s$ increases, but their sequence is not reversed [Fig.~\ref{fig:effective_mass_large_rs}(a)]. Thus the decreasing behavior of $m_\mathrm{H}^*(\theta)$ observed at small $r_s<2$ persists with increasing $r_s$ beyond $r_s>2$ as explicitly shown in Fig.~\ref{fig:effective_mass_large_rs}(c). 
For $m_\mathrm{L}^*$, however, the effective mass curves cross each other at around $r_s\sim7$, completely reversing their sequence at $r_s>8$ as compared to the case for $r_s<6$. This leads $m_\mathrm{L}^*(\theta)$ to have qualitatively different behaviors from $m_\mathrm{H}^*(\theta)$ as shown in Fig.~\ref{fig:effective_mass_large_rs}(d). The effective mass $m_\mathrm{L}^*(\theta)$ shows a decreasing behavior before the crossover occurs [$r_s=5.0$ in Fig.~\ref{fig:effective_mass_large_rs}(d)]. When one enters the crossover regime ($r_s=7.0$), the effective mass at small $\theta$ starts increasing, yielding a local maximum off the symmetry axes. The local maximum is shifted to the right with increasing $r_s$, expanding the region where the effective mass increases. For $r_s$ beyond the crossover regime, the effective mass monotonically increases over the whole range of $\theta$ ($r_s=10.0$). Note that for these larger values of $r_s$ used in Fig.~\ref{fig:effective_mass_large_rs}, the RPA-$GW$ theory becomes progressively quantitatively worse, but it is known that even for metals with $r_s\sim6$, the $GW$ theory provides reasonable results although the perturbation expansion of the inset in Fig.~\ref{fig:FS_schematic} is no longer valid for large $r_s$ \cite{Hedin1965,Rice1965}. This could be because the effective expansion parameter at large $r_s$ may be renormalized to an effectively much smaller value as has been argued theoretically \cite{Zhang2005, Zhang2005a}.

{\em Fermi Surface.}---
Using Eq.~(\ref{eq:renorm_energy}), we can obtain the renormalized Fermi surface by solving
\begin{equation}
    \mu^*=\frac{k_{\mathrm{F}x}^2}{2m_\mathrm{H}^*(\bm k_\mathrm{F})}+\frac{k_{\mathrm{F}y}^2}{2m_\mathrm{L}^*(\bm k_\mathrm{F})},
\end{equation}
where $\mu^*$ is the renormalized Fermi energy. Assuming that $m_\mathrm{H}^*(\bm k)$ and $m_\mathrm{L}^*(\bm k)$ vary slower than $k_x^2$ and $k_y^2$ near the Fermi surface, respectively, we obtain
\begin{align}
    k_\mathrm{F}(\theta)=\frac{ \sqrt{2\mu^*}\sqrt{m_\mathrm{H}^*(\theta) m_\mathrm{L}^*(\theta)} }{\sqrt{ m_\mathrm{H}^*(\theta)\sin^2{\theta} + m_\mathrm{L}^*(\theta)\cos^2{\theta}  }},
    \label{eq:Fermi_surface}
\end{align}
where we parametrize the Fermi surface by $\bm k_\mathrm{F} = k_\mathrm{F}(\theta)\bm{\hat{k}}_\mathrm{F}$. Here $\theta$ is the angle from the axis corresponding to the high mass direction. To describe the deviation of the Fermi surface from an ellipse, we define 
\begin{equation}
\eta(\theta)=\frac{k_\mathrm{F}(\theta)}{\tilde{k}_\mathrm{F}(\theta)},    
\label{eq:eta}
\end{equation}
where $\tilde{k}_\mathrm{F}(\theta)$ represents the Fermi surface approximated as an ellipse and thus is given by the standard equation of an ellipse, with $k_\mathrm{F}(\theta=0)$ and $k_\mathrm{F}(\theta=\pi/2)$ being the semi-major and semi-minor axes, respectively:
\begin{align}
    \tilde{k}_\mathrm{F}(\theta)=\frac{ \sqrt{2\mu^*}\sqrt{m_\mathrm{H}^*(0) m_\mathrm{L}^*\left(\frac{\pi}{2}\right)} }{\sqrt{ m_\mathrm{H}^*(0)\sin^2{\theta} + m_\mathrm{L}^*\left(\frac{\pi}{2}\right)\cos^2{\theta}  }}.
    \label{eq:Fermi_surface_ellipse}
\end{align}
Note that $\mu^*$ drops out of Eq.~(\ref{eq:eta}), and thus we need only the effective mass to describe the deviation of the Fermi surface from an ellipse. 

For exact results, Eq.~(\ref{eq:Fermi_surface}) should be solved in a self-consistent manner because the effective mass is evaluated at the renormalized Fermi surface. This requires obtaining the effective mass from the self-consistent Dyson equation [Eq.~(\ref{eq:Dyson})], i.e., within the off-shell approximation. It is clear that the off-shell approximation is exact if we work with the full self-energy. But it has been argued that because of vertex corrections, the on-shell approximation is the appropriate approximation to be used in the $GW$ theory so that different perturbative orders are not mixed in the results \cite{DuBois1959,DuBois1959a,Rice1965,Lee1975,Ting1975,Zhang2005,Zhang2005a}. Thus we use our on-shell effective mass results presented in the previous section for the calculation of the Fermi surface.

\begin{figure}[!htb]
    \centering
    \includegraphics[width=1\linewidth]{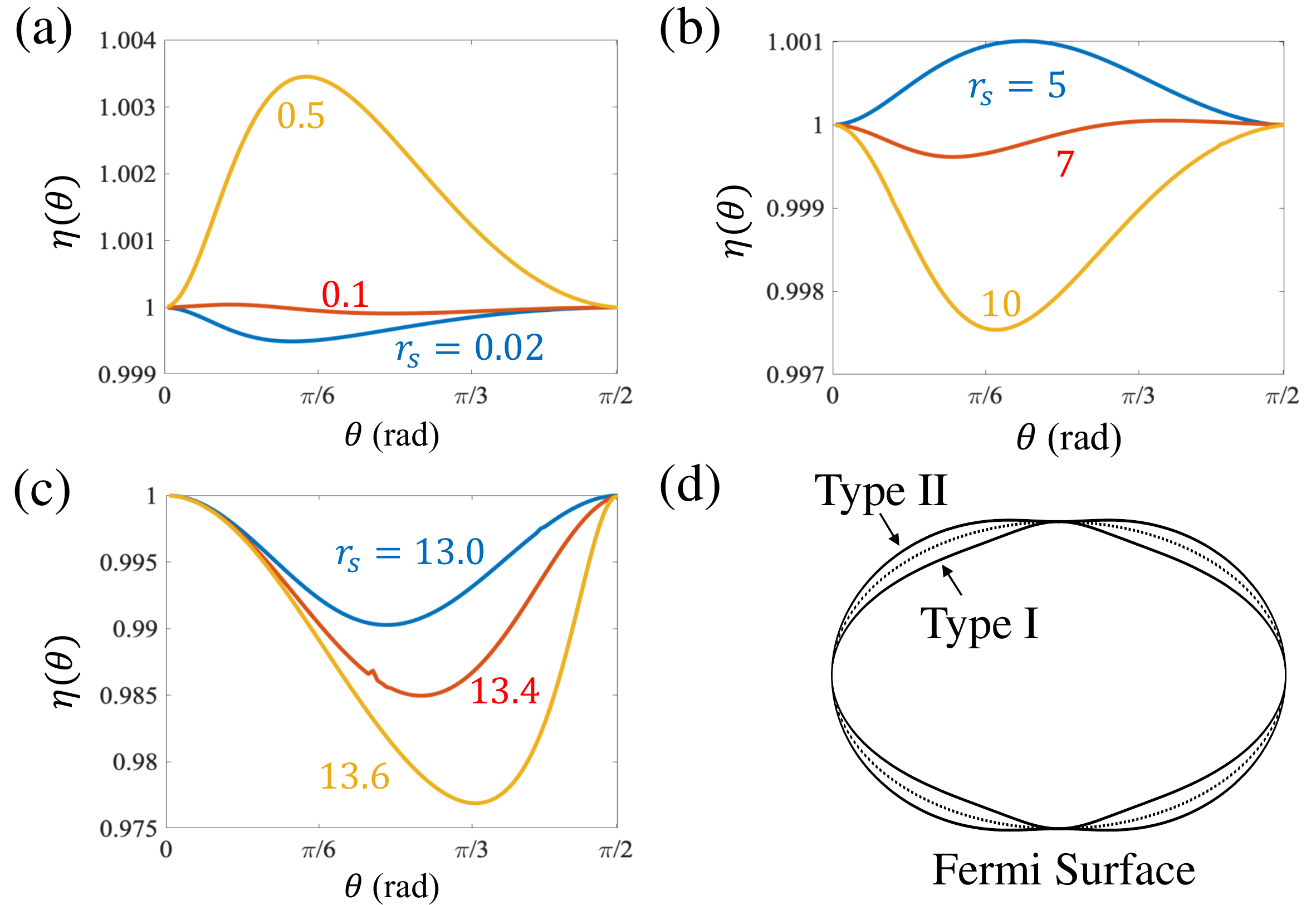}
    \caption{Plots of $\eta(\theta)$ in (a), (b) crossover regimes and (c) for larger $r_s$. (d) The shape of Fermi surfaces corresponding to a convex (type I) and concave (type II) shape of $\eta(\theta)$ along with the ellipse that approximates the renormalized Fermi surfaces}
    \label{fig:Fermi_surface}
\end{figure}

In the previous section, we have shown that there are two crossover regimes (one around $r_s\sim0.1$ and the other $r_s\sim8$) where $m^*(\theta)$ changes its behavior qualitatively. In the following we investigate its consequence on the Fermi surface. Figure~\ref{fig:Fermi_surface}(a) presents $\eta(\theta)$ in the first crossover regime around $r_s\sim0.1$. At very small $r_s=0.02$, $\eta(\theta)$ exhibits a convex shape, for which the corresponding Fermi surface shape is schematically shown in Fig.~\ref{fig:Fermi_surface}(d) as type I. When $r_s$ increases to $0.1$, $\eta(\theta)$ becomes almost flat, and the corresponding Fermi surface is close to an ideal ellipse. As we increase $r_s$ further, $\eta(\theta)$ becomes concave, and the corresponding Fermi surface becomes type II in Fig.~\ref{fig:Fermi_surface}(d). Thus, the Fermi surface qualitatively changes its shape from type I to type II in the first crossover regime. Until $r_s$ increases up to the second crossover regime, the shape of the Fermi surface does not qualitatively change because there is no qualitative change in the angular behavior of the effective mass. In the second crossover regime, a similar but opposite behavior is observed: $\eta(\theta)$ undergoes a concave to convex transition [see Fig~\ref{fig:Fermi_surface}(b)], and thus the Fermi surface changes its shape from type II to type I.

After the second (last) crossover regime, the Fermi surface maintains its type I shape as $r_s$ increases further. We show in Fig.~\ref{fig:Fermi_surface}(c) plots of $\eta(\theta)$ for still larger $r_s$. Since the effective mass increases more rapidly at larger $r_s$ [see the inset in Fig.~\ref{fig:effective_mass_large_rs}(b)], even a small increase of $r_s$ for large $r_s$ leads to a substantial change of $\eta(\theta)$. Whether this interesting interaction-driven qualitative topology change in the Fermi surface at large $r_s$ is real or an artifact of our RPA approximation is unknown at this time and should be experimentally investigated in gated 2D systems by varying electron density so as to change $r_s$ \cite{Ando1982, Zhang2005, Zhang2005a, Ting1975, Lee1975,Shashkin2002}. We note that the theory \cite{Zhang2005a} predicts an effective mass divergence in the isotropic system at $r_s\sim16$, which was experimentally observed \cite{Shashkin2002}. 

\begin{figure}[tb]
    \centering
    \includegraphics[width=1\linewidth]{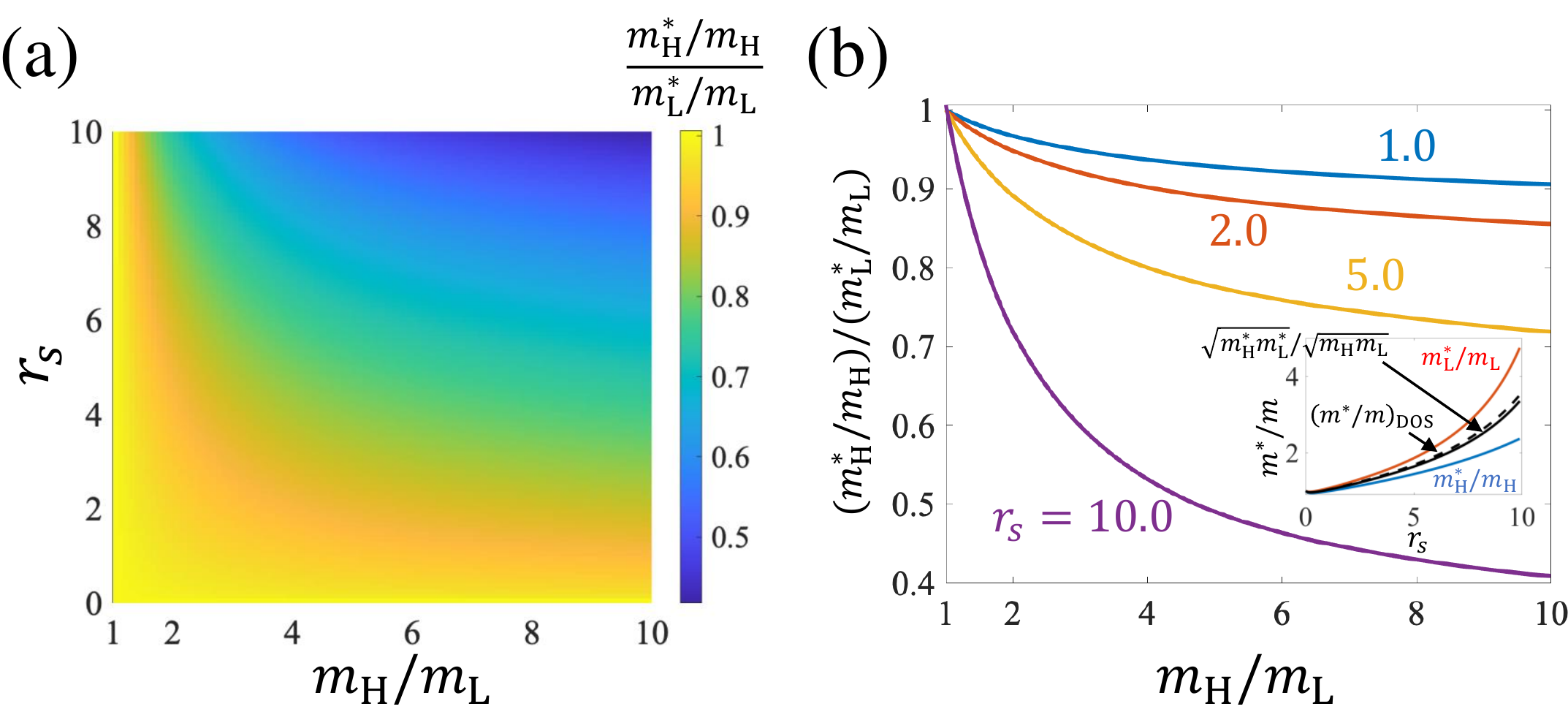}
    \caption{(a) Calculated RPA-$GW$ effective mass anisotropy renormalization as a function of bare anisotropy ($m_\mathrm{H}/m_\mathrm{L}$) and the interaction strength $r_s$—the blue indicates strong suppression of the anisotropy by interaction; (b) line plots for several $r_s$ values corresponding to panel (a)—note that for large $r_s$, the bare anisotropy is strongly suppressed.  The inset shows the anisotropy as well as the density of states mass in the presence of interaction for a fixed $m_\mathrm{H}/m_\mathrm{L}$=5. Here $(m^*/m)_\mathrm{DOS}$ (black dashed line) refers to a result for an isotropic electron gas with the density of states mass.  }
    \label{fig:renorm_mass}
\end{figure}

Since the predicted nonelliptic corrections of our theory are small, except for large $r_s$ where the RPA may not be a good approximation, we also investigate the interaction effect on the mass anisotropy itself assuming an elliptic approximation, which remains an excellent approximation for $r_s<10$.  These results of `anisotropy renormalization' are shown in Fig.~\ref{fig:renorm_mass}, and the conclusion is obvious:  The bare anisotropy is suppressed strongly by interaction effects with the renormalized $m^*_\mathrm{H}/m^*_\mathrm{L}$ being much less than the bare $m_\mathrm{H}/m_\mathrm{L}$. Our work thus provides somewhat of a justification for the widespread use of the isotropic approximation in the interacting many body system, using the renormalized density of states effective mass, even when the corresponding bare system has strong anisotropy.

{\em Summary.}---
Within the highly successful leading order dynamically screened RPA theory, we have studied the distortion of the Fermi surface by Coulomb interactions in an anisotropic two-dimensional electron gas.
We find that the distorted Fermi surface has only two qualitatively distinct shapes, which we classify as type I and type II. A transition between the two shapes can occur as $r_s$ varies, but only in a limited range of $r_s$.
Our predictions 
can be experimentally tested in gated 2D semiconductor structures by varying $r_s$ through varying carrier density. Our predicted anisotropic Fermi surface effects, although rather small quantitatively, should show up in experiments in anisotropic 2D electron systems, e.g. Si 111 and Si 110 inversion layers \cite{Ando1982} and 2D AlAs layers \cite{Chung2018}, in any measurement directly involving the effective mass such as specific heat, Shubnikov–de Haas (SdH) oscillations, cyclotron resonance, and electrical transport. We also calculate the interaction-induced renormalization of the mass anisotropy within an elliptic approximation, showing that interaction strongly suppresses the bare effective mass anisotropy.

\begin{acknowledgments}
This work is supported by the Laboratory for Physical Sciences.
\end{acknowledgments}

\bibliography{ref}
\end{document}